\documentclass[aps,prl,reprint,groupedaddress]{revtex4-1}
\usepackage{amssymb}
\usepackage{amsmath}
\usepackage{array}
\usepackage{multirow,bigdelim}
\usepackage{enumitem}

\begin{document}

\title{ Reducing the detection of genuine entanglement of n qubits to two
qubits}
\author{Dafa Li$^{1,2}$}

\begin{abstract}
We propose a criterion for the detection of genuine entanglement of pure
multiqubit states. To this aim, we define an operator called the losing one
qubit operator, which is different from the reduced density operator. The
states obtained from a multiqubit state by applying the losing one qubit
operator are referred to as its projected states. We show that all of the
projected states of a pure product n-qubit state are pure product states
provided that it cannot be written as a product of a single qubit state and
a genuinely entangled (n-1)-qubit state. We also show that a pure n-qubit
state is genuinely entangled provided that the state has at least two
genuinely entangled (n-1)-qubit projected states. By repeating the losing
process, we reduce the detection of entanglement of pure n-qubit states to
the one of pure two-qubit states. Also we write a LISP program for the
reduction process.
\end{abstract}

\affiliation{
$^1$Department of Mathematical Sciences, Tsinghua University,
Beijing, 100084, China\\
$^2$Center for Quantum Information Science and Technology, Tsinghua National
Laboratory for Information Science and Technology (TNList), Beijing,
100084, China\\}


\maketitle

\section{Introduction}

Quantum entanglement is considered as a unique quantum mechanical resource
\cite{Nielsen}. It is well known that entanglement takes a key rule in
quantum information processing tasks, for e.g., quantum teleportation,
quantum cryptography and quantum key distribution.

It is known that for GHZ of three qubits, tracing out of qubit i, the
reduced density operator becomes completely unentangled, while for W of
three qubits, tracing out of qubit i the reduced density operator remains
entangled \cite{Dur}. It is indicated that many physical implementations of
qubits, for example ion traps, optical lattices and linear optics, suffer
from loss of qubits \cite{Wallman}. The entanglement resistant to particles
loss via tracing out the particles was investigated \cite{Quinta}.

Many efforts have been devoted to exploring criteria for detection of
quantum entanglement \cite{Peres}-\cite{Osterloh}.A necessary condition for
separability of a quantum system consisting of two subsystems is that a
matrix, obtained by partial transposition of $\rho $, has only non-negative
eigenvalues \cite{Peres}. \ The criterion is referred to as PPT (Positive
partial transpose). For $2\times \ 2$ and $2\times \ 3$ systems, the
positivity of the partial transposition of a state is necessary and
sufficient for its separability \cite{Horodecki}.

It was shown that for any separable state of a bipartite system, the sum of
the singular values of the realigned matrix constructed from the density
matrix is necessarily not greater than 1 \cite{Chen}. The generalized
reduction criterion for the separability of bipartite system in arbitrary
dimensions was proposed \cite{Albeverio}. The cross norm necessary criterion
for the separability of density matrices for bipartite systems was studied
\cite{Rudolph}.

Recently, Zwerger et al. showed that genuine entanglement of all
multipartite pure states can be detected in a device-independent way via
bipartite Bell inequalities \cite{Zwerger}. The genuine $n$-qubit
entanglement can be detected via the proportionality of two coefficient
vectors \cite{DLI-QIP-19}.

In this paper, we propose the losing one qubit operator. Applying the losing
one qubit operator to a state of $n$ qubits, we can obtain $n$ states of $%
(n-1)$ qubits which are referred to as the projected states. We demonstrate
that the losing one qubit operator can reduce the detection of genuine
entanglement of n qubits to two qubits. In section 2, we give necessary and
sufficient conditions for two, three, and four qubits to be product states.
In section 3, we show that all of the projected states of an$\ n$-qubit
product state are product states provided that the $n$-qubit product state
cannot be written as a product of a single qubit state and a genuinely
entangled $(n-1)$-qubit state. Thus, the losing one qubit operator can
reduce the detection of the entanglement of n-qubit states to $(n-1)$-qubit
ones and finally to two-qubit ones.

\ \ \ \ \ \ \ \ \ \ \ \ \ \ \ \

\section{The necessary and sufficient conditions for product states of two,
three, and four qubits}

For given two vectors $v_{1}$, and $v_{2}$, if $v_{1}=kv_{2}$, where $k$ is
a complex number, then we say that $v_{1}$ is proportional to $v_{2}$.
Specially, when $v_{1}=0$, then $k=0$. For a set of vectors $v_{1}$, $v_{2}$,%
$\cdots $, and $v_{m}$, if there is a vector $v_{i}\neq 0$ of the set of
vectors such that any vector $v_{\ell }$ of the set of vectors is
proportional to $v_{i}$, then we say that the set of vectors are
proportional. Clearly, if at least two non-zero vectors of the set of the
vectors are not proportional, then clearly, the set of the vectors are not
proportional.

\subsection{Two qubits,}

For two qubits, we can write any pure state of two qubits as $|\psi \rangle
_{12}=\sum_{i=0}^{3}c_{i}|i\rangle $.

\textit{Result 1}. It is well known that $|\psi \rangle _{12}$ is a product
state iff the following two vectors are proportional \cite{DLI-QIP-19}

\begin{equation}
\left(
\begin{array}{cc}
c_{0} & c_{1}%
\end{array}%
\right) ^{T},\left(
\begin{array}{cc}
c_{2} & c_{3}%
\end{array}%
\right) ^{T}.
\end{equation}

Note that the above two vectors are proportional iff the following equality
holds.
\begin{equation}
c_{0}c_{3}=c_{1}c_{2}.  \label{conc}
\end{equation}%
Note also that $|c_{0}c_{3}-c_{1}c_{2}|^{2}$ is just the concurrence of $%
|\psi \rangle _{12}$.

\subsection{Three qubits}

\textit{Result 2.} Let $|\psi \rangle _{123}=\sum_{i=0}^{7}c_{i}|i\rangle $
be any pure state of three qubits. Then, $|\psi \rangle _{123}$ is a product
state iff at least one of the following three pairs of vectors is
proportional. Otherwise, it is genuinely entangled \cite{DLI-QIP-19}.

\begin{equation}
\left(
\begin{array}{cccc}
c_{0} & c_{1} & c_{2} & c_{3}%
\end{array}%
\right) ^{T},\left(
\begin{array}{cccc}
c_{4} & c_{5} & c_{6} & c_{7}%
\end{array}%
\right) ^{T}
\end{equation}

\begin{equation}
\left(
\begin{array}{cccc}
c_{0} & c_{1} & c_{4} & c_{5}%
\end{array}%
\right) ^{T},\left(
\begin{array}{cccc}
c_{2} & c_{3} & c_{6} & c_{7}%
\end{array}%
\right) ^{T}
\end{equation}

\begin{equation}
\left(
\begin{array}{cccc}
c_{0} & c_{2} & c_{4} & c_{6}%
\end{array}%
\right) ^{T},\left(
\begin{array}{cccc}
c_{1} & c_{3} & c_{5} & c_{7}%
\end{array}%
\right) ^{T}
\end{equation}

\subsection{Four qubits}

Result 3. Let $|\psi \rangle _{1234}=\sum_{i=0}^{15}c_{i}|i\rangle $. A pure
product state $|\psi \rangle $ of four qubits can be written as (1). $%
|\varphi \rangle _{1}|\phi \rangle _{234}$, (2). $|\varphi \rangle _{2}|\phi
\rangle _{134}$, (3). $|\varphi \rangle _{3}|\phi \rangle _{124}$, (4). $%
|\varphi \rangle _{4}|\phi \rangle _{123}$, (5). $|\varphi \rangle
_{12}|\phi \rangle _{34}$, (6). $|\varphi \rangle _{13}|\phi \rangle _{24}$,
or (7). $|\varphi \rangle _{14}|\phi \rangle _{23}$ iff the following
corresponding set of vectors are proportional.

\begin{eqnarray}
&&\text{(1).}\left(
\begin{array}{cccccccc}
c_{0} & c_{1} & c_{2} & c_{3} & c_{4} & c_{5} & c_{6} & c_{7}%
\end{array}%
\right) ^{T},  \notag \\
&&\left(
\begin{array}{cccccccc}
c_{8} & c_{9} & c_{10} & c_{11} & c_{12} & c_{13} & c_{14} & c_{15}%
\end{array}%
\right) ^{T};  \label{vec-1}
\end{eqnarray}

\begin{eqnarray}
&&\text{(2). }\left(
\begin{array}{cccccccc}
c_{0} & c_{1} & c_{2} & c_{3} & c_{8} & c_{9} & c_{10} & c_{11}%
\end{array}%
\right) ^{T},  \notag \\
&&\left(
\begin{array}{cccccccc}
c_{4} & c_{5} & c_{6} & c_{7} & c_{12} & c_{13} & c_{14} & c_{15}%
\end{array}%
\right) ^{T};  \label{vec-2}
\end{eqnarray}

\begin{eqnarray}
&&\text{(3). }\left(
\begin{array}{cccccccc}
c_{0} & c_{1} & c_{4} & c_{5} & c_{8} & c_{9} & c_{12} & c_{13}%
\end{array}%
\right) ^{T},  \notag \\
&&\left(
\begin{array}{cccccccc}
c_{2} & c_{3} & c_{6} & c_{7} & c_{10} & c_{11} & c_{14} & c_{15}%
\end{array}%
\right) ^{T};  \label{vec-3}
\end{eqnarray}

\begin{eqnarray}
&&\text{(4). }\left(
\begin{array}{cccccccc}
c_{0} & c_{2} & c_{4} & c_{6} & c_{8} & c_{10} & c_{12} & c_{14}%
\end{array}%
\right) ^{T},  \notag \\
&&\left(
\begin{array}{cccccccc}
c_{1} & c_{3} & c_{5} & c_{7} & c_{9} & c_{11} & c_{13} & c_{15}%
\end{array}%
\right) ^{T};
\end{eqnarray}
\
\begin{eqnarray}
&&\text{(5). }\left(
\begin{array}{cccc}
c_{0} & c_{1} & c_{2} & c_{3}%
\end{array}%
\right) ^{T},\left(
\begin{array}{cccc}
c_{4} & c_{5} & c_{6} & c_{7}%
\end{array}%
\right) ^{T},  \notag \\
&&\left(
\begin{array}{cccc}
c_{8} & c_{9} & c_{10} & c_{11}%
\end{array}%
\right) ^{T},\left(
\begin{array}{cccc}
c_{12} & c_{13} & c_{14} & c_{15}%
\end{array}%
\right) ^{T};
\end{eqnarray}

\begin{eqnarray}
&&\text{(6). }\left(
\begin{array}{cccc}
c_{0} & c_{1} & c_{4} & c_{5}%
\end{array}%
\right) ^{T},\left(
\begin{array}{cccc}
c_{2} & c_{3} & c_{6} & c_{7}%
\end{array}%
\right) ^{T},  \notag \\
&&\left(
\begin{array}{cccc}
c_{8} & c_{9} & c_{12} & c_{13}%
\end{array}%
\right) ^{T},\left(
\begin{array}{cccc}
c_{10} & c_{11} & c_{14} & c_{15}%
\end{array}%
\right) ^{T};  \label{vec-6}
\end{eqnarray}

\begin{eqnarray}
&&\text{(7). }\left(
\begin{array}{cccc}
c_{0} & c_{1} & c_{8} & c_{9}%
\end{array}%
\right) ^{T},\left(
\begin{array}{cccc}
c_{2} & c_{3} & c_{10} & c_{11}%
\end{array}%
\right) ^{T},  \notag \\
&&\left(
\begin{array}{cccc}
c_{4} & c_{5} & c_{12} & c_{13}%
\end{array}%
\right) ^{T},\left(
\begin{array}{cccc}
c_{6} & c_{7} & c_{14} & c_{15}%
\end{array}%
\right) ^{T}.  \label{vec-7}
\end{eqnarray}

Proof. We only prove Case (1): a pure product state $|\psi \rangle
=\sum_{i=0}^{15}c_{i}|i\rangle $ of four qubits can be written as $|\varphi
\rangle _{1}|\phi \rangle _{234}$ iff the the set of vectors in Eq. (\ref%
{vec-1}) are proportional.

($\Longrightarrow $). Let $|\psi \rangle =|\varphi \rangle _{1}|\phi \rangle
_{234}$, where $|\varphi \rangle _{1}=(\alpha |0\rangle _{1}+\beta |1\rangle
_{1})$ and $|\phi \rangle _{234}=\sum_{i=0}^{7}a_{i}|i\rangle _{234}$.
Clearly, $|\psi \rangle =$ $\alpha \sum_{i=0}^{7}a_{i}|0\rangle
_{1}|i\rangle _{234}+\beta \sum_{i=0}^{7}a_{i}|1\rangle _{1}|i\rangle _{234}$%
. Let $v_{1}$ be the first vector and $v_{2}$ be the second vector in Eq. (%
\ref{vec-1}). Then,
\begin{eqnarray}
v_{1} &=&\left(
\begin{array}{cccccccc}
\alpha a_{0} & \alpha a_{1} & \alpha a_{2} & \alpha a_{3} & \alpha a_{4} &
\alpha a_{5} & \alpha a_{6} & \alpha a_{7}%
\end{array}%
\right) ^{T},  \notag \\
v_{2} &=&\left(
\begin{array}{cccccccc}
\beta a_{0} & \beta a_{1} & \beta a_{2} & \beta a_{3} & \beta a_{4} & \beta
a_{5} & \beta a_{6} & \beta a_{7}%
\end{array}%
\right) ^{T}.
\end{eqnarray}

Thus, one can see that $v_{1}$ and $v_{2}$ are proportional.

($\Longleftarrow $). Conversely, if $v_{1}$ and $v_{2}$ are proportional,
then we can write $v_{2}=kv_{1}$ or $v_{1}=kv_{2}$. Assume that $%
v_{2}=kv_{1} $ and $v_{1}=\left(
\begin{array}{cccc}
c_{0} & c_{1} & \cdots & c_{7}%
\end{array}%
\right) ^{T}$. Then, via $v_{1}$ and $v_{2}$\ we can write

\begin{eqnarray}
|\psi \rangle &=&\sum_{i=0}^{7}c_{i}|0\rangle _{1}|i\rangle
_{234}+k\sum_{i=0}^{7}c_{i}|1\rangle _{1}|i\rangle _{234}  \label{tensor-2}
\\
&=&|\varphi \rangle _{1}|\phi \rangle _{234},
\end{eqnarray}%
where $|\varphi \rangle _{1}=$ $(|0\rangle _{1}+k|1\rangle _{1})$ and $|\phi
\rangle _{234}=\sum_{i=0}^{7}c_{i}|i\rangle _{234}$.

Thus, to determine that a state of four qubits is genuinely entangled, we
need to determine that each of above seven sets of vectors are not
proportional. We investigated the separability of four qubits by using
permutations of qubits \cite{DLI-QIP-19}.

\section{Losing one qubit operator}

\subsection{Losing one qubit operator for three qubits}

First we use the following example to demonstrate the losing one qubit
operator. We consider a product state of three qubits, for e.g., $|\psi
\rangle _{123}=|\phi \rangle _{2}|\varphi \rangle _{13}$, where $|\phi
\rangle _{2}=(\alpha |0\rangle _{2}+\beta |1\rangle _{2})$ and $|\varphi
\rangle _{13}=(a|00\rangle _{13}+b|01\rangle _{13}+c|10\rangle
_{13}+d|11\rangle _{13})$.

There are the following cases for losing qubits.

Losing qubit 1:

Let $|\psi ^{\ast }\rangle _{123/1}=_{1}\langle 0|\psi \rangle
_{123}+_{1}\langle 1|\psi \rangle _{123}$. Then, a calculation yields that

$|\psi ^{\ast }\rangle _{123/1}=(\alpha |0\rangle _{2}+\beta |1\rangle
_{2})((a+c)|0\rangle _{3}+(b+d)|1\rangle _{3})$. Note that if the
coefficients of $|\psi ^{\ast }\rangle _{123/1}$ vanish, then we also call $%
|\psi ^{\ast }\rangle _{123/1}$ a product state. Thus, clearly $|\psi ^{\ast
}\rangle _{123/1}$ is a product state.

Losing qubit 2:

Let $|\psi ^{\ast }\rangle _{123/2}=_{2}\langle 0|\psi \rangle
_{123}+_{2}\langle 1|\psi \rangle _{123}$. Then, a calculation yields that $%
|\psi ^{\ast }\rangle _{123/2}=(\alpha +\beta )|\varphi \rangle _{13}$.
Then, $|\psi ^{\ast }\rangle _{123/2}$ is an entangled state iff $|\varphi
\rangle _{13}$ is an entangled state whenever $\alpha +\beta \neq 0$.

Losing qubit 3:

Let $|\psi ^{\ast }\rangle _{123/3}=$\ $_{3}\langle 0|\psi \rangle
_{123}+_{3}\langle 1|\psi \rangle _{123}$. Then, a calculation yields that $%
|\psi ^{\ast }\rangle _{123/3}=$ $(a+b)(\alpha |00\rangle _{12}+\beta
|01\rangle _{12})+(c+d)(\alpha |10\rangle _{12}+\beta |11\rangle _{12})$. It
is easy to see that the concurrence vanishes for $|\psi ^{\ast }\rangle
_{123/3}$. Via Eq. (\ref{conc}), therefore $|\psi ^{\ast }\rangle _{123/3}$
is a product state.

So, for a product state of three qubits, there is at most one qubit $i$, $%
i=1,2,$ or $3$, such that $|\psi ^{\ast }\rangle _{123/i}$ is entangled.

A calculation yields the Table 1.

Table 1.

\begin{tabular}{|c|c|c|c|}
\hline
3-qubits & $|\psi ^{\ast }\rangle _{123/1}$ & $|\psi ^{\ast }\rangle
_{123/2} $ & $|\psi ^{\ast }\rangle _{123/3}$ \\ \hline
$|000\rangle $ & product & product & product \\ \hline
$|0\rangle _{1}|EPR\rangle _{23}$ & entangled & product & product \\ \hline
GHZ & entangled & entangled & entangled \\ \hline
W & entangled & entangled & entangled \\ \hline
\end{tabular}

\subsection{Reducing the detection of the entanglement of n qubits to (n-1)
qubits}

Definition: Let $|\psi \rangle _{1\cdots n}$ be a state of $n(\geq 3)$
qubits and $|\psi ^{\ast }\rangle _{1\cdots n/i}=_{i}\langle 0|\psi \rangle
_{1\cdots n}+_{i}\langle 1|\psi \rangle _{1\cdots n}$. We call the $(n-1)$%
-qubit state $|\psi ^{\ast }\rangle _{1\cdots n/i}$ a projected state of $%
|\psi \rangle _{1\cdots n}$ obtained by losing qubit $i$.

We can conclude the following theorem.

Theorem 1. If $|\psi \rangle _{1\cdots n}$ is a product state of $n$ qubits,
then there exists at most one qubit $i$ such that $|\psi ^{\ast }\rangle
_{1\cdots n/i}$ is genuinely entangled. That is, for $j\neq i$, $|\psi
^{\ast }\rangle _{1\cdots n/j}$ are product states or zero.

Proof. It is trivial for $n=2$. For $n\geq 3$ qubits, the proof is put in
Appendix A.

Corollary 1. Let $|\psi \rangle _{1\cdots n}$ be a state of $n$ qubits. If
there are at least two qubits $i$ and $j$ such that $(n-1)$-qubit projected
states $|\psi ^{\ast }\rangle _{1\cdots n/i}$ and $|\psi ^{\ast }\rangle
_{1\cdots n/j}$\ are genuinely entangled, then the $n$-qubit state $|\psi
\rangle _{1\cdots n}$ is genuinely entangled.

We give the following example to show that Corollary 1 is not necessary.

Let $|\psi \rangle _{123}=|001\rangle +|010\rangle +|100\rangle +|111\rangle
$ (belonging to W SLOCC\ class). Then, $|\psi ^{\ast }\rangle _{123/1}=|\psi
^{\ast }\rangle _{123/2}=|\psi ^{\ast }\rangle _{123/3}=(|0\rangle
+|1\rangle )(|0\rangle +|1\rangle )$. Though $|\psi \rangle _{123}$ is
genuinely entangled, $|\psi ^{\ast }\rangle _{123/i}$, $i=1,2,3$, are
product states.

Corollary 2. If $|\psi \rangle _{1\cdots n}$ is an $n$-qubit product state
but not as \ a product of a single qubit state and an ($n-1$)-qubit
genuinely entangled state, then the projected state $|\psi ^{\ast }\rangle
_{1\cdots n/i}$ is a product state for any qubit $i$.

Proof. Ref. the proof of Theorem 1.

From Corollary 2, it is easy to show that the following Corollary 3 is true.

Corollary 3. If $|\psi \rangle _{1\cdots n}$, which is not as a product of a
single qubit state and an $(n-1)$-qubit genuinely entangled state, has a
genuinely entangled $(n-1)$-qubit projected state, then $|\psi \rangle
_{1\cdots n}$ is genuinely entangled.

Via Corollary 1, for three qubits, we derive a simple sufficient condition
for genuinely entangled states in Appendix B.

\subsection{Reducing the detection of the entanglement of n qubits to two
qubits}

For any pure state $|\psi \rangle _{1\cdots
n}=\sum_{i=0}^{2^{n}-1}a_{i}|i\rangle $, in light of Corollary 1,\ if there
are at least two $(n-1)$-qubit projected states\ which are genuinely
entangled, then the $n$-qubit state $|\psi \rangle _{1\cdots n}$ is
genuinely entangled. For the $(n-1)$-qubit projected state $|\psi ^{\ast
}\rangle _{1\cdots n/i}$, in light of Corollary 1, if it has at least two $%
(n-2)$-qubit projected states\ which are genuinely entangled, then the $%
(n-1) $-qubit state $|\psi ^{\ast }\rangle _{1\cdots n/i}$ is genuinely
entangled. The process repeats until we get $4$ (3, or 2)-qubit projected
states. Therefore, detecting genuine entanglement of n qubits may reduce to
four, three, or two qubits.

For four, three and two qubits, we have given the simple necessary and
sufficient conditions for detecting genuine entanglement in Result 1, Result
2 and Result 3.

Example 1. For the $n$-qubit W state, all the ($n-1$) qubit projected states
are of the form $\sqrt{\frac{n-1}{n}}$W$_{n-1}+\frac{1}{\sqrt{n}}|0\cdots
0\rangle $, where W$_{n-1}$ is the $(n-1)$-qubit W state. \ The $k$-qubit
projected states are of the form $\sqrt{\frac{k}{n}}$W$_{k}+\frac{n-k}{\sqrt{%
n}}|0\cdots 0\rangle $. Specially, 2-qubit projected states have the form $%
\sqrt{\frac{2}{n}}(|01\rangle +|10\rangle )+\frac{n-2}{\sqrt{n}}|00\rangle $%
. The latter state has nonzero concurrence and therefore is entangled.
Consequently, via Corollary 1, the n-qubit W state is genuinely entangled.

Example 2. Let the $n$-qubit state $|\psi \rangle _{1\cdots n}=$ $\alpha
|i_{1}i_{2}\cdots i_{n}\rangle +\beta |\overline{i_{1}}\overline{i_{2}}%
\cdots \overline{i_{n}}\rangle $, where $\alpha \beta \neq 0$, $i_{l}=0$ or
1 and $\overline{i_{l}}=1-i_{l}$. Clearly, after losing one qubit\ $k$, we
obtain the projected state $|\psi ^{\ast }\rangle _{1\cdots n/k}=\alpha
|i_{1}\cdots i_{(k-1)}i_{(k+1)}\cdots i_{n}\rangle +\beta |\overline{i_{1}}%
\cdots \overline{i_{(k-1)}}\overline{i_{(k+1)}}\cdots \overline{i_{n}}%
\rangle $. We can continue applying the losing one qubit operator to $|\psi
^{\ast }\rangle _{1\cdots n/k}$. Finally, we obtain two-qubit projected
states $\alpha |z_{1}z_{2}\rangle +\beta |\overline{z_{1}}\overline{z_{2}}%
\rangle $. It is easy to know that $\alpha |z_{1}z_{2}\rangle +\beta |%
\overline{z_{1}}\overline{z_{2}}\rangle $ is entangled. Therefore $|\psi
\rangle _{1\cdots n}$ is genuinely entangled via Corollary 1. Specially, the
$n$-qubit GHZ is genuinely\ entangled.

Example 3. Let us check that $|\psi \rangle _{1234}=|0000\rangle
+|0111\rangle -|1111\rangle $ is genuinely entangled. A calculation produces
that $|\psi ^{\ast }\rangle _{1234/1}=|000\rangle $, which is a product
state, and $|\psi ^{\ast }\rangle _{1234/i}=|000\rangle +|011\rangle
-|111\rangle $, $i=2,3,4$. Next, we show that $|\psi ^{\ast }\rangle
_{1234/i}$, $i=2,3,4$, are genuinely entangled. Let $|\omega \rangle
_{123}=|000\rangle +|011\rangle -|111\rangle $. A calculation yields that $%
|\omega ^{\ast }\rangle _{123/1}=$ $|00\rangle $ and $|\omega ^{\ast
}\rangle _{123/2}=|\omega ^{\ast }\rangle _{123/3}=|00\rangle +|01\rangle
-|11\rangle $, which is an entangled state of two qubits. So, via Corollary
1, the three-qubit states $|\psi ^{\ast }\rangle _{1234/i}$, $i=2,3,4$, are
genuinely entangled, and then $|\psi \rangle _{1234}$ is genuinely\
entangled.

\subsection{Maximally entangled states (MES)}

MES can be defined via different ways such as entropy and LOCC. It is well
known that the GHZ state can be regarded as the maximally entangled state of
three qubits in several aspects.

Let $|\psi \rangle _{1\cdots n}$ be a state of $n$ qubits. If $k$ of all the
$(n-1)$-qubit projected states $|\psi ^{\ast }\rangle _{1\cdots n/i}$\ are
genuinely entangled, then we say that $|\psi \rangle _{1\cdots n}$ has the
entanglement measure of $k$. If all the $(n-1)$-qubit projected states are
genuinely entangled, then the state is called MES.

We next demonstrate the entanglement measure of some entangled states below.

(1). After losing one qubit, the projected states of the $n$-qubit GHZ state
are just the $(n-1)$-qubit GHZ state. It means that losing one qubit
operator preserves the GHZ-constructions. Clearly, GHZ is a MES.

(2). For the W state of $n$ qubits, all the $(n-1)$-qubit projected states
are $|P\rangle =$ $\sqrt{\frac{n-1}{n}}$W$_{n-1}+\frac{1}{\sqrt{n}}$ $%
|0\cdots 0\rangle $, which is not the W-states. It means that losing one
qubit operator does not preserve the W-construction. Thus, the
W-construction is fragile under losing one qubit operator. From Example 1, $%
|P\rangle $ is genuinely\ entangled. Therefore, W is a MES.

(3).\ $|\Phi _{4}\rangle =|0001\rangle +|0010\rangle +|1100\rangle
+|1111\rangle $ is genuinely entangled \cite{Osterloh}. $|\Phi _{4}^{\ast
}\rangle _{1234/4}=|\Phi _{4}^{\ast }\rangle _{1234/3}=(|00\rangle
+|11\rangle )(|0\rangle +|1\rangle )$ which are product states while $|\Phi
_{4}^{\ast }\rangle _{1234/2}=|\psi ^{\ast }\rangle _{1234/1}=|001\rangle
+|010\rangle +|100\rangle +|111\rangle $ which are genuinely\ entangled.
Thus, $|\Phi _{4}\rangle $ has the\ entanglement measure 2.

\subsection{A program for losing one qubit operator}

We write a LISP program for the formula for the projected states $|\psi
^{\ast }\rangle _{1\cdots n/i}$ and the procedure reducing detection of
entanglement of n qubits to two qubits in Appendix C. The program can detect
the genuinely entangled states with the measure $\geq 2$. For example, the
program can detect the following genuinely\ entangled states.

(1) The states GHZ and W of three qubits

(2). For four qubits, GHZ, W, Cluster state, Dicke state $|2,4\rangle $, and
Osterloh's $|\Phi _{5}\rangle $ and $|\Phi _{4}\rangle $ states \cite%
{Osterloh}.

(3). For five qubits, Osterloh's $|\Psi _{2}\rangle ,|\Psi _{5}\rangle ,$
and $|\Psi _{6}\rangle $ states \cite{Osterloh}.

(4). For six qubits, Osterloh's $|\Xi _{2}\rangle ,|\Xi _{5}\rangle ,|\Xi
_{6}\rangle ,$ and $|\Xi _{7}\rangle $ states \cite{Osterloh}.

\subsection{Comparing reduced density operator with losing one qubit operator%
}

(1). Reduced density operator and losing one qubit operator are different.

For the Bell state $|$Bell$\rangle =\frac{1}{\sqrt{2}}(|00\rangle
+|11\rangle )$, the reduced density is $tr_{2}(|$Bell$\rangle \langle $Bell$%
|)=\frac{1}{2}I$. After losing one qubit, $|\psi ^{\ast }\rangle _{12/2}=%
\frac{1}{\sqrt{2}}(|0\rangle +|1\rangle )$. Therefore, losing a qubit
operator is different from the reducing density operator.

(2). $|\psi ^{\ast }\rangle _{1\cdots n/i}$ is a pure state while the
reducing density $tr_{i}(|\psi \rangle _{1\cdots n}\langle \psi |)$ may
become a mixed state. For example, $tr_{2}(|$Bell$\rangle \langle $Bell$|)=%
\frac{1}{2}I$, which is a mixed state.

(3). After losing one qubit, the $(n-1)$-qubit projected states of the $n$%
-qubit GHZ state are still the $(n-1)$-qubit GHZ state while for GHZ of
three qubits, tracing out of a qubit, the reduced density operator becomes
completely unentangled. It means that the entanglement property of the state
GHZ is fragile under tracing out a qubit. Ref. Table 2.

Table 2.

\begin{tabular}{|c|c|c|}
\hline
3 qubits & $\rho ^{12},\rho ^{23},\rho ^{13}$ & $|\psi ^{\ast }\rangle
_{123/i},i=1,2,3$ \\ \hline
GHZ & separable & entangled, GHZ \\ \hline
W & entangled & entangled, not W \\ \hline
\end{tabular}

\section{Summary}

In this paper, we show that all the projected states obtained via the losing
one qubit operator are product states for a product state of $n$ qubits but
not as a product of a single qubit state and a genuinely entangled $(n-1)$%
-qubit state. Thus, if there are two $(n-1)$-qubit projected states\ which
are genuinely entangled, then the state of $n$ qubit is genuinely entangled.
We can repeat the process until we get $2$ (3,or 4)-qubit projected states
for which we have the necessary and sufficient conditions to detect their
separability. Thus, the losing one qubit operator can reduce the detection
of entanglement for $n$-qubits to two qubits. We have written a LISP program
for detection of entanglement for $n$-qubits.

\section{Appendix A The proof of Theorem 1}

The proof of Theorem is follows.

For $n\geq 3$ qubits, there are three cases.

Case 1. $|\psi \rangle _{1\cdots n}=|\phi \rangle |\varphi \rangle |\omega
\rangle $. Clearly, it is easy to see that $|\psi ^{\ast }\rangle _{12\cdots
n/i}$ is a product state for any $i$.

Case 2. $|\psi \rangle _{1\cdots n}=|\phi \rangle _{q_{1}\cdots
q_{i}}|\varphi \rangle _{q_{i+1}\cdots q_{n}}$, where $i\geq 2$, $n-i\geq 2$.

Clearly, $|\psi \rangle _{1\cdots n/q_{k}}^{\ast }=$ $|\phi \rangle
_{q_{1}\cdots q_{i}/q_{k}}^{\ast }|\varphi \rangle _{q_{i+1}\cdots q_{n}}$,
where $1\leq k\leq i$. One can see that $|\psi \rangle _{1\cdots
n/q_{k}}^{\ast }$ is a product state of $(n-1)$ qubits or zero.

Case 3. $|\psi \rangle _{1\cdots n}=(\alpha |0\rangle _{q_{1}}+\beta
|1\rangle _{q_{1}})|\varphi \rangle _{q_{2}\cdots q_{n}}$, where $|\varphi
\rangle _{q_{2}\cdots q_{n}}$ is genuinely\ entangled. For example, $|\psi
\rangle _{123}=|0\rangle _{1}|$EPR$\rangle _{23}$, $|\psi ^{\ast }\rangle
_{123/1}=|$EPR$\rangle _{23}$.

Case 3.1. $|\psi ^{\ast }\rangle _{12\cdots n/q_{1}}=(\alpha +\beta
)|\varphi \rangle _{q_{2}\cdots q_{n}}$, which is genuinely entangled when $%
\alpha +\beta \neq 0$ or zero when $\alpha +\beta =0$.

Case 3.2. $|\varphi \rangle _{q_{2}\cdots q_{n}}=|0\rangle _{q_{k}}|\varphi
^{(0)}\rangle _{q_{2}\cdots q_{n}/q_{k}}+|1\rangle _{q_{k}}|\varphi
^{(1)}\rangle _{q_{2}\cdots q_{n}/q_{k}}$, $k\neq 1$.

Thus, $|\psi ^{\ast }\rangle _{12\cdots n/q_{k}}=(\alpha |0\rangle
_{q_{1}}+\beta |1\rangle _{q_{1}})|\varphi ^{\ast }\rangle _{q_{2}\cdots
q_{n}/q_{k}}$ which is a product state.

From the above three cases, we can conclude If $|\psi \rangle _{1\cdots n}$
is a product state of $n$ qubits, then there exists at most one qubit $q_{i}$
such that $|\psi ^{\ast }\rangle _{12\cdots n/q_{i}}$ is genuinely
entangled. That is, for other $q_{j}$, $|\psi ^{\ast }\rangle _{12\cdots
n/q_{j}}$ are product states.

\section{Appendix B. A simple sufficient condition for genuinely entangled
states of three qubits}

Via Corollary 1, for three qubits, we derive a simple sufficient condition
for genuinely\ entangled states by losing one qubit operator. Let $|\psi
\rangle _{123}=\sum_{i=0}^{7}c_{i}|i\rangle $ be any pure state of three
qubits.

A calculation yields that
\begin{eqnarray*}
&&|\psi ^{\ast }\rangle _{123/1} \\
&=&(c_{0}+c_{4})|00\rangle +(c_{1}+c_{5})|01\rangle \\
&&+(c_{2}+c_{6})|10\rangle +(c_{3}+c_{7})|11\rangle .
\end{eqnarray*}

Via Eq. (\ref{conc}), clearly, $|\psi ^{\ast }\rangle _{123/1}$ is entangled
iff
\begin{equation}
(c_{0}+c_{4})(c_{3}+c_{7})\neq (c_{1}+c_{5})(c_{2}+c_{6}).  \label{con-1}
\end{equation}

A calculation yields that
\begin{eqnarray*}
&&|\psi ^{\ast }\rangle _{123/2} \\
&=&(c_{0}+c_{2})|00\rangle +(c_{1}+c_{3})|01\rangle \\
&&+(c_{4}+c_{6})|10\rangle +(c_{5}+c_{7})|11\rangle .
\end{eqnarray*}

Via Eq. (\ref{conc}), $|\psi ^{\ast }\rangle _{123/2}$ is entangled iff
\begin{equation}
(c_{0}+c_{2})(c_{5}+c_{7})\neq (c_{1}+c_{3})(c_{4}+c_{6}).  \label{con-2}
\end{equation}

A calculation yields that%
\begin{eqnarray*}
&&|\psi ^{\ast }\rangle _{123/3} \\
&=&(c_{0}+c_{1})|00\rangle +(c_{2}+c_{3})|01\rangle \\
&&+(c_{4}+c_{5})|10\rangle +(c_{6}+c_{7})|11\rangle .
\end{eqnarray*}

Via Eq. (\ref{conc}), $|\psi ^{\ast }\rangle _{123/3}$ is entangled iff
\begin{equation}
(c_{0}+c_{1})(c_{6}+c_{7})\neq (c_{2}+c_{3})(c_{4}+c_{5}).  \label{con-3}
\end{equation}

Via Corollary 1, $|\psi \rangle _{123}=\sum_{i=0}^{7}c_{i}|i\rangle $ is
genuinely entangled if at least two of Eqs. (\ref{con-1}, \ref{con-2}, \ref%
{con-3}) hold.

\section{Appendix C A formula for the projected states}

Let $|\psi \rangle _{1\cdots n}=\sum_{i=0}^{2^{n}-1}a_{i}|i\rangle $ be any
pure state of n qubits. Then, we can write $|\psi \rangle _{1\cdots
n}=|0\rangle _{k}|\psi ^{(0)}\rangle _{1\cdots n/k}+|1\rangle _{k}|\psi
^{(1)}\rangle _{1\cdots n/k}$. Therefore, $|\psi ^{\ast }\rangle _{1,\cdots
,n/k}=|\psi ^{(0)}\rangle _{1\cdots n/k}+|\psi ^{(1)}\rangle _{1\cdots n/k}$%
. Denoting $\ell =2^{n-k}$, then $|\psi ^{\ast }\rangle _{1\cdots n/k}$

$=\sum_{i=0}^{\ell -1}(a_{0\times \ell +i}+a_{1\times \ell +i})|\underbrace{%
0\cdots 0}_{k-1}\rangle |i\rangle $

$+\sum_{i=0}^{\ell -1}(a_{2\times \ell +i}+a_{3\times \ell +i})|\underbrace{%
0\cdots 01}_{k-1}\rangle |i\rangle +\cdots $

$+\sum_{i=0}^{\ell -1}(a_{(2^{k}-4)\times \ell +i}+a_{(2^{k}-3)\times \ell
+i})|\underbrace{1\cdots 10}_{k-1}\rangle |i\rangle $

$+\sum_{i=0}^{\ell -1}(a_{(2^{k}-2)\times \ell +i}+a_{(2^{k}-1)\times \ell
+i})|\underbrace{1\cdots 10}_{k-1}\rangle |i\rangle $.

\end{document}